# Understanding Compositional Evolution of Hollows at Dominici Crater, Mercury

A. Emran*,[1] and K. M. Stack[1]

[1] NASA Jet Propulsion Laboratory, California Institute of Technology, Pasadena, CA 91109, USA

*Highlights:*

- Hollow spectra suggest silicates with lower sulfide and graphite concentrations.
- Hollows form through the thermal decomposition and destabilization of sulfides and graphite leading to loss of sulfur and carbon.
- Mixture of silicates remain primarily as lag deposits after material loss.

**Abstract**

Hollows on Mercury are small depressions formed by volatile loss, providing important clues about the volatile inventory of the planet's surface and shallow subsurface. We investigate the composition of hollows in various phases of devolatilization at Dominici crater. By applying a machine learning approach to MESSENGER Mercury Dual Imaging System data, we defined surface units within the study area and extracted their reflectance spectra. We applied linear (areal) spectral modeling using laboratory sulfides, chlorides, graphite, and silicate mineral spectra to estimate the composition of hollows and their surrounding terrains. At Dominici, the hollow on the crater rim/wall is interpreted to be active, while that in the center of the crater is interpreted as a waning hollow. We find that the active hollow predominantly comprises silicates (augite and albite), with a trace amount of graphite and CaS. In contrast, waning hollows contain marginally elevated sulfides (MgS and CaS) and graphite, but slightly lower silicates than the active hollow. The spectra of low reflectance terrain surrounding the hollows appear to be dominated by graphite and sulfides, which contribute to its darker appearance. We suggest that hollow at the crater forms due to thermal decomposition of sulfides, primarily MgS possibly mixed with CaS, as well as possible the depletion of graphite. As devolatilization wanes, a mixture of predominantly silicate minerals remains in the hollows — impeding further vertical growth.

*Corresponding author: A. Emran (al.emran@jpl.nasa.gov).

1. **Introduction**

Hollows are shallow (tens of meters), flat-floored, irregularly shaped, and rimless depressions on Mercury with relatively high-reflectance interiors that are sometimes encircled by bright halos (Blewett et al. 2011, 2013; Thomas et al. 2014). Hollows have diameters of tens of meters to several kilometers (Blewett et al. 2011, 2018) and can occur in isolation or in clusters (Thomas et al. 2014). Hollows are typically found in the walls, rims, ejecta, central peaks, and floors of impact crater structures and multi-ring basins and occur across the planet (Blewett et al. 2011, 2013; Thomas et al. 2014). These unique geologic features are considered the youngest and freshest non-impact landforms on the planet, and they may still be actively forming today (e.g., Blewett et al. 2013, 2018; Thomas et al. 2014). While hollows are not exclusive to any surface unit, they preferentially occur in and associated with low reflectance materials (LRM; Blewett et al. 2011, 2013; Thomas et al. 2014, 2016), a globally extensive surface unit on Mercury (Denevi et al. 2009) with visible to near-infrared (VIS-NIR) spectral reflectance and slope lower than the planetary average (Robinson et al. 2008; Blewett et al. 2009; Murchie et al. 2015).

A sequence for the development and evolution of hollows has been proposed in which a hollow is initiated with the appearance of a small bright depression within a "dark spot", a feature of the LRM that has relatively lower reflectance than the surrounding terrains (Xiao et al. 2013). Several factors and processes are thought to contribute to hollow formation, including thermal desorption by solar heating, solar UV radiation and wind-stimulated desorption, photon bombardment, chemical sputtering from magnetospheric ions, pyroclastic volcanism, and vaporization from micrometeorite impact (e.g., Blewett et al. 2013, 2016, 2018; Thomas et al. 2014, 2016). As a hollow continues to grow and enlarge in diameter by likely scarp retreat and coalescence of nearby hollows, the dark spot disappears and is replaced by the formation of a bright halo (Xiao et al. 2013; Blewett et al. 2013, 2018). As volatiles loss, remnant material is left behind as lag deposits, and the bright halo fades until the brightness of the hollow matches that of the surrounding terrains (Xiao et al. 2013; Blewett et al. 2018). Environmental factors, such as topographic steep slopes, can influence hollow growth (Deutsch et al. 2025). Steep slopes favor mass wasting, which exposes volatile-bearing species at or near the surface, making them more susceptible to thermal instability due harsh environment of the planet (e.g., Blewett et al. 2016; Vilas et al. 2018). Note that since hollows are not exclusively formed within "dark spot" in LRM (Blewett et al. 2013), the formation scenario outlined above might not be the only possible mechanism (Xiao et al. 2013). Hollows exhibiting high-reflectance interiors and halos are interpreted as actively forming (Blewett et al. 2011, 2013, 2018), while hollows with dark-toned interiors and exteriors similar to surrounding terrains have been interpreted as inactive (Blewett et al. 2011, 2013). If a hollow has only a high reflectance interior, it is considered to be in a waning stage of evolution (Blewett et al. 2011, 2013).

Though the volatile-bearing species (thermally unstable at/near surface exposure or susceptible to decomposition) has been identified as a potential cause of hollow formation (Blewett et al. 2016, 2018; Thomas et al. 2014, 2016; Wang et al. 2020; Barraud et al. 2023), a complete understanding of the development and evolution of hollows on the planet remains elusive. Outstanding questions include: What significant volatiles and non-volatiles occur within LRM and Mercury's surface and subsurface? Which species take part in sublimation process and responsible for hollow formation? Which mineral phases are responsible for constraining further growth of hollow development? Further assessment of hollow composition during different phases of devolatilization is, therefore, warranted to understand the origin of these features and the composition of the planet's interior and surface. Constraining the formation and composition of hollows at different development stages has the potential to advance our understanding of Mercury's geological history, bulk mineralogy, and volatile inventory (e.g., Thomas et al. 2016; Wang et al. 2020; Lucchetti et al. 2021).

In this study, we used a machine learning approach to delineate and extract MErcury Surface, Space ENvironment, GEochemistry, and Ranging (MESSENGER) Mercury Dual Imaging System (MDIS) instrument (Hawkins et al. 2007) spectra from hollows at different development stages within Dominici crater, an ~20 km diameter rayed crater in the Kuiper quadrangle (Blewett et al. 2013; Lucchetti et al. 2018), to understand the evolution of hollows at different devolatilization phases (Fig. 1). The hollow at the crater rim has both high-reflectance interiors and halos, whereas the hollow at the crater center has a high-reflectance interior but lacks a surrounding halo (e.g., Vilas et al. 2016) as observed in higher-resolution images from the MDIS instrument onboard MESSENGER (Fig. 2a). Thus, the hollows in Dominici crater are in both active and waning stages— based on the interpretation of Blewett et al. (2011, 2013 and 2018)— and its hollows exhibit distinct spectral characteristics as reported in previous studies (e.g., Vilas et al. 2016;

Lucchetti et al. 2018). These characteristics enable a comprehensive understanding of hollow composition without spatial or observational bias, despite the planet's diverse mineralogy (e.g., Namur and Charlier, 2017; McCoy et al. 2018; Nittler and Weider, 2019).

## 2. Observations and Methods

### 2.1. Dataset and calibration

The MDIS instrument (Hawkins et al. 2007) consists of two cameras: a multispectral Wide-Angle Camera (WAC) with a 10.5° x 10.5° field of view and a monochrome higher resolution Narrow Angle Camera (NAC) with a 1.5° x 1.5° field of view (Hawkins et al. 2007, 2009). We used MDIS-WAC multispectral observations with eleven spectral channels between 433.2 and 1012.6 nm (0.43 – 1.01 μm). The details of the MDIS experimental data records (EDR) products used in this study are provided in Table A1. Images were corrected for radiometric calibration with a standard routine and photometrically calibrated to a standard geometry of an incidence angle $i = 30°$, emission angle $e = 0°$, and phase angle $g = 30°$ using the Hapke model (Domingue et al. 2015, 2016). We included an empirical correction routine (Keller et al. 2013) in the processing pipeline to adjust the responsivity correction of the MDIS image filter (Denevi et al. 2017). Images were projected to an equirectangular projection and layer stacked to make a multispectral image mosaic at a spatial resolution of 326 m/pixel. Standard calibration routines were performed using the United States Geological Survey (USGS)'s Integrated Software for Imagers and Spectrometers (ISISv3) package (Adoram-Kershner et al. 2020).

Both the spatial and spectral resolution of the MDIS-WAC images used in this study is higher than the data used in Vilas et al. (2016) and Lucchetti et al. (2018), enabling new insights into the composition and characteristics of the Dominici hollows. While Vilas et al. (2016) and Lucchetti et al. (2018) used MDIS images with eight spectral bands at ~ 1000 m/pixel resolution, this study used MDIS images with eleven spectral channels with ~3x times higher spatial resolution (326 m/pixel). We did not use higher spectral resolution data from the Mercury Atmospheric and Surface Composition Spectrometer (MASCS)'s Visible and Infrared Spectrograph (VIRS; McClintock and Lankton, 2007) instrument because the different hollows at the crater are below the spatial scale of the MASCS-VIRS footprint. Moreover, the characteristic broad absorption 0.558- 0.828 µm in hollow spectra reported from MDIS-WAC data (Vilas et al. 2016; Lucchetti et al. 2018) has not been confidently observed in VIRS spectra (Izenberg et al. 2014; Murchie et al. 2015; Thomas et al. 2016; Barraud et al. 2020, 2023).

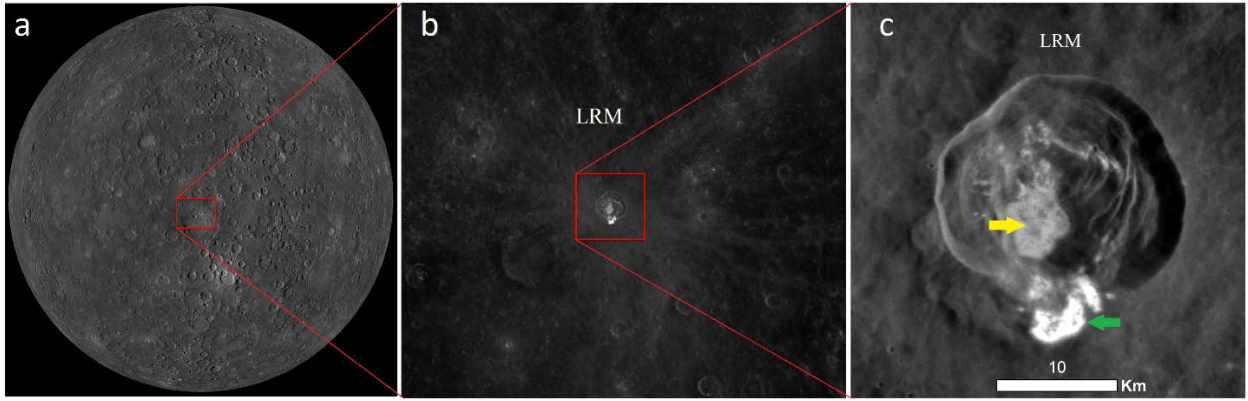

**Fig. 1**: Locational context of hollows at Dominici crater with central coordinates at 1.26ºN, 323.5ºE. (a) Location of the crater on Mercury in the MDIS (Hawkins et al. 2007) basemap. (b) Regional context of the crater in the MDIS-WAC filter at 0.699 µm. (c) Hollows at the crater in higher resolution MDIS-NAC observation at 0.749 µm band. The yellow arrow points to the lower reflectance waning stage hollow, while the green arrow points to the brighter active hollow with a halo. The LRM is labeled in the subplots (b) and (c). North is up.

## 2.2. Spectral mapping

We implemented principal component-reduced Gaussian mixture model (PC-GMM; Emran et al. 2023a) to the MDIS-WAC calibrated image mosaic to define surface spectral units within and surrounding the crater. PC-GMM combines principal component analysis (PCA; e.g., Abdi and Williams, 2010) to reduce the original data dimension, followed by the application of the multivariate Gaussian mixture model (GMM; e.g., Bishop, 2006) to identify distinct clusters in the reduced data. The number of principal components (PC-axes) was selected automatically using the hyperspectral signal identification by minimum error algorithm (Bioucas-Dias and Nascimento, 2008; Appendix A2; ref. Fig. A1). The Akaike information criterion (AIC; Akaike, 1974) and Bayesian information criterion (BIC; Schwarz, 1978) were used to select the optimal number of clusters for GMM (e.g., Emran et al. 2023a, 2023b; Appendix A2; ref. Fig. A2). We use Python modules for PCA and GMM (Pedregosa et al. 2011) to process the calibrated MDIS-WAC data. Details of the PC-GMM used in this study are provided in the Appendix A2. The use of PC-GMM and higher spatial and spectral resolution MDIS-WAC data in this study offered advantages over K-mean clustering (Marzo et al. 2006) employed previously by Lucchetti et al. (2018) (Emran et al. 2023a and reference therein). Particularly, GMM has the advantage of uncovering complex patterns and identifying heterogeneity in the data, making the resulting clusters more representative of real-world scenario (e.g., Patel & Kushwaha, 2020; Emran et al. 2023a). While K-mean clustering (Marzo et al. 2006) has been successful in many instances, it has certain limitations, as it relies on rigid segmentation, which may not effectively capture the inherent heterogeneity of real-world data (e.g., Patel & Kushwaha, 2020; Emran et al. 2023a). Furthermore, while K-mean partitions data into non-overlapping groups, GMM can infer the probability of overlapping clusters, better reflecting the complexity of real-world datasets (e.g., Patel & Kushwaha, 2020; Emran et al. 2023a). The application of PC-GMM has been shown successfully in analyzing Pluto's surface composition at both global and local scales (Emran et al. 2023a, 2023b), and also holds potential for application to other planetary bodies in the Solar System.

Seven distinct spectral classes/clusters were identified at Dominici crater (Fig. 2). For convenience, we hereafter denote a cluster by 'C' followed by the assigned class number (e.g., C1, C2, C3, etc.). The probability on a scale of 0 to 1, where a higher value represents a greater likelihood, of each surface spectral class occurring at each MDIS-WAC pixel is provided in Fig. A3. The average reflectance and standard deviation ("uncertainty") spectra of each spectral class were extracted from the class's pixels for further investigation and comparison of their compositions. We calculated the normalized average spectra at 0.558 μm to compare results with Lucchetti et al. (2018) and produced continuum-removed spectra using the convex hull approach (e.g., Clark and King, 1987) to detect weak (subtle) spectral absorption bands for all surface units (Fig. 2).

We found that the standard deviations of the C3 spectra were several times higher than those of the other clusters. These higher values could be attributed to relatively higher noise, topographic influence, or variable compositions. This outcome was anticipated, as we classified the area using denoised (refined) data and then extracted the corresponding reflectance at WAC channels from the original MDIS data. Thus, we further processed the C3 spectra such that the spectral mean remained the same while the standard deviation was reduced (Sciamma-O'Brien et al. 2025; Diane H. Wooden, personal communication). This additional processing does not alter the characteristic average reflectance spectra; instead, it refined the standard deviation for C3, providing better constraints for the spectral modeling used in the following section. This approach first calculated a "processing factor" for each C3 pixel by dividing the sum of the pixel's reflectance across all wavelengths by the sum of the cluster's average reflectance over the same wavelengths. A "processed reflectance" at each pixel was then calculated by dividing the reflectance of the pixel by the processing factor of the pixel. The reduced standard deviation of C3 was calculated from the processed reflectance of all pixels at this cluster. For reference, the average spectra and standard deviation of each cluster are provided in Fig. A4.

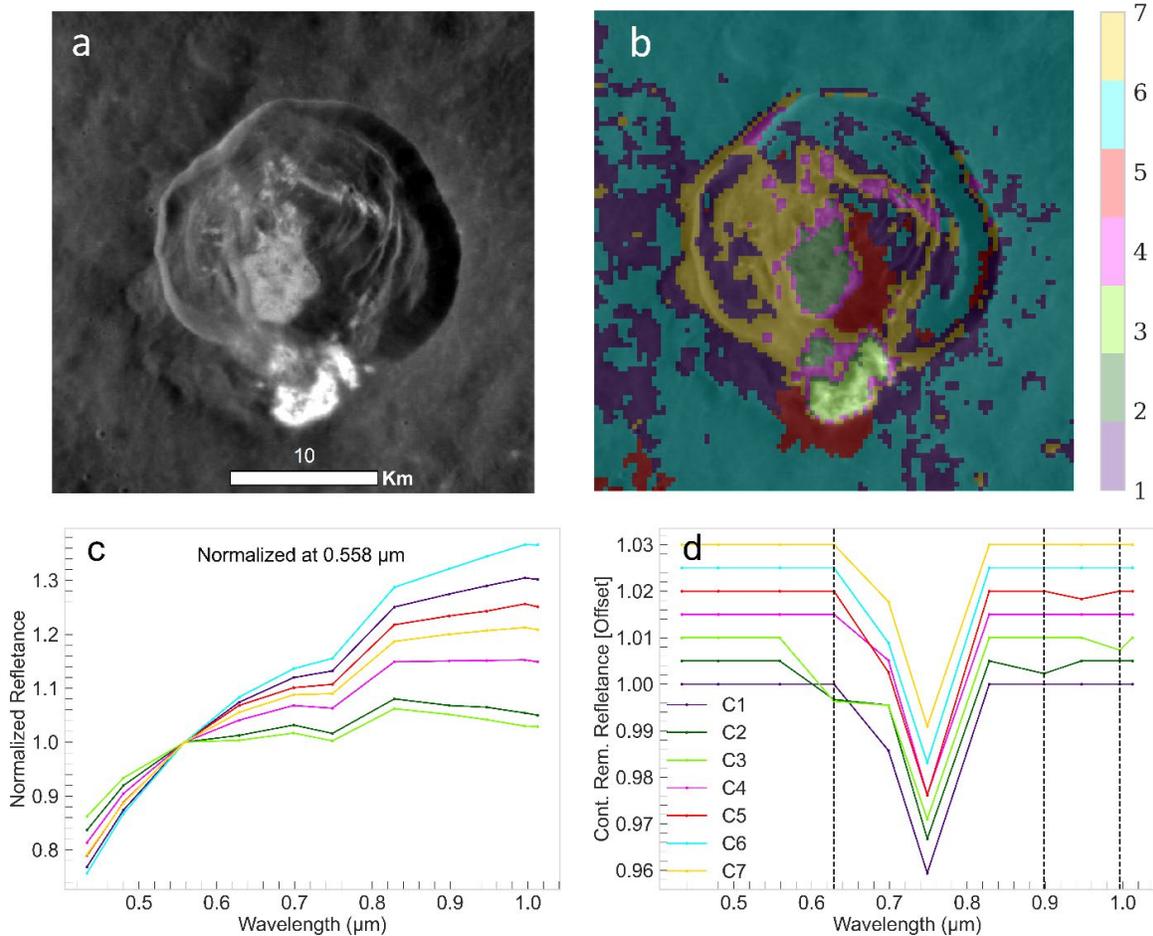

**Fig. 2**: (a) Panchromatic observation of the crater and surrounding areas in targeted MDIS-NAC data (CN0253965560M) at 50 m/pixel. (b) Seven distinct spectral classes were identified using PC-GMM (Emran et al. 2023a, 2023b) at the crater and surrounding terrain in the background MDIS-NAC image. (c) Normalized reflectance spectra at 0.558-μm for each surface unit. (d) Continuum-removed spectra using the convex hull approach (e.g., Clark and King, 1987) for seven spectral classes, each showing the largest absorption feature at 0.75 μm (a shoulder at ~0.7 μm) with a similar strength across all units. The dashed vertical lines indicate absorption features at ~0.63, 0.9, and 1.0 μm. The colors of the spectra correspond to the colors in the cluster map in (b).

## 2.3. Spectral ratio

The reflectance spectra from hollow surface units were compared to laboratory reflectance spectra of sulfides (CaS, MgS, CrS, FeS, MnS, $Na_2S$, and $TiS_2$), chlorides ($CaCl_2$, $MgCl_2$, and NaCl), graphite, and silicate minerals such as pyroxene (enstatite, augite, and diopside) and plagioclase (labradorite and albite). We chose these laboratory endmembers because they are representative of Mercury's analog mineral phases and the volatile-bearing species associated with hollows (e.g., Blewett et al. 2016, 2018; Vander Kaaden and McCubbin, 2015; Lucchetti et al. 2016, 2020; Namur and Charlier, 2017; Wang et al. 2020; Barraud et al. 2023). The reflectance spectra of sulfides, both fresh and heated samples with a grain size of ~10 μm, were measured at a phase angle of 26º– as prepared by Varatharajan et al. (2019). The reflectance spectra of chlorides, graphite, and silicate minerals were measured at a phase angle of 30º and were collected from the RELAB database (Milliken et al. 2021). The grain sizes of RELAB spectra vary from 0-250 μm for chloride, 0-45 μm for graphite, 45-90 μm for pyroxenes, and 0-25 μm for plagioclase samples (Table A2). For graphite, we considered three different samples and averaged these spectra. All laboratory spectra were resampled to MDIS-WAC wavelengths. Details of the RELAB reflectance spectra of laboratory endmembers used in this study are given in Table A2. Note that the phase angles of all laboratory spectra are comparable to MDIS-WAC observations (i.e. 30º).

We investigated spectral slopes at visible and infrared wavelength within MDIS-WAC data for the Dominici and laboratory spectra. This was accomplished by comparing the ratio of 430 nm to 750 nm reflectance (VISr) and 950 nm to 750 nm reflectance (IRr) for the average spectra of each cluster and the laboratory reflectance spectra of candidate endmembers (Fig. 3). The absorption band at 0.75 μm is unlikely related to volatile phase (e.g., Wang et al. 2020) so we used the reflectance at this wavelength as the denominator for ratioing reflectance spectra both for VISr and IRr while comparing the observational data and laboratory spectral endmembers. A comparison of the VISr and IRr spectral indices was used to select the appropriate laboratory endmembers for spectral modeling outlined in the next section. Moreover, analysis of the visible spectral slope (VISr) can directly compare results of the same index from other hollows using MDIS-WAC data (Blewett et al. 2013; Thomas et al. 2016), and a similar spectral index of 415 nm/750 nm (Izenberg et al. 2014, 2015) and VIS spectral slope 445-750 nm (Barraud et al. 2020) from the MASCS-VIRS (McClintock and Lankton, 2007) onboard MESSENGER.

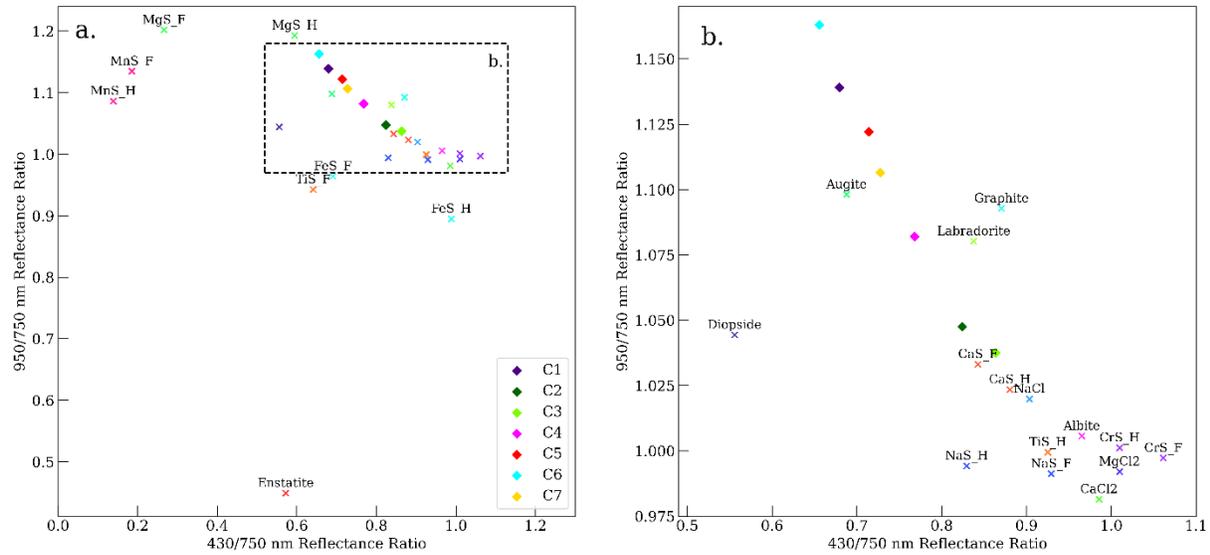

**Fig. 3**: Comparison of the spectral ratios of seven spectral units at Dominici crater and surrounding terrain to the reflectance spectra of laboratory endmembers. (a) Spectral ratio of VISr 430 nm/750 nm reflectance (x-axis) and IRr 950 nm/750 nm reflectance (y-axis). Diamonds are the clusters at the crater and surrounding areas- the color of each diamond corresponds to the colors in the clustering map in Fig. 2. The C2 (dark green) and C3 (lawn green) diamonds represent the waning and active hollows, respectively, at the crater. Cross symbols represent the laboratory endmembers. For sulfides, *_F and *_H represent fresh and heated samples, respectively. (b) Zoomed-in version of the subplot in (a).

## *2.4. Spectral modeling*

We applied a linear (areal) mixing model to estimate the fractional contribution of mineral endmembers to each surface unit. Although the linear mixing model has both advantages and limitations (e.g., Keshava and Mustard, 2002; Hapke, 2012), we used linear mixing approach because investigations of Mercury's surface composition have been successfully conducted using linear models with both MDIS-WAC data (Lucchetti et al. 2021) and MASCS-VIRS (Barraud et al. 2023) data from the MESSENGER spacecraft. Laboratory endmembers for spectral modeling were selected based on the results from the spectral ratio analysis (*Section 2.3*). Since all seven surface units exhibited 430 nm/750 nm (VISr) and 950 nm/750 nm (IRr) values less than and greater than 1, respectively (Fig. 3), we chose laboratory endmembers that met these spectral ratio criteria. Therefore, the reflectance spectra of both fresh and heated samples of sulfides (CaS, MgS, and MnS), NaCl, graphite, augite, diopside, labradorite, and albite were used as the laboratory endmembers for the spectral model (Fig. 4). The linear mixing model assumes an areal mineral mixture, where reflectance is modeled using a linear mixture of endmembers spectra (e.g., Stack and Milliken, 2015; Emran et al. 2021; Lucchetti et al. 2021; Barraud et al. 2023). According to the linear model, the reflectance of a mixture consists of the sum of the spectra of individual components weighted by their corresponding fractional contribution such that:

$$R_{(\lambda)} = C \sum_{i=0}^{k} f_i * r_{i,\lambda} \qquad (1)$$

$$0 \leq f_i \leq 1 \qquad (2)$$

$$\sum_{i=0}^{k} f_i = 1 \qquad (3)$$

where, where $r_{i,\lambda}$ is the reflectance of the $i^{th}$ endmember at $\lambda$ wavelength and $f_i$ is the fraction contributed by the $i^{th}$ endmember, and $C$ is a scaling factor. The scaling factor can account for small variations between observational and laboratory data (e.g., Davis & Brown, 2024), which may arise due to factors such as slight differences in viewing geometry between the measurements. To estimate the fractional contribution of constituents for each surface unit, we employed the non-negative constrained least squares (NNLS; Bro and De Jong, 1997) method. The NNLS algorithm adopts a non-negativity constraint in the model fit, ensuring non-negative contributions from each laboratory endmember, as negative mineral contributions are not physically realistic.

Sulfides exhibit changes in spectral characteristics at visible wavelengths upon thermal processing (Helbert et al. 2013). These changes are most pronounced for MgS, while CaS and MnS show much subtler variations (Fig. 4; Helbert et al. 2013; Varatharajan et al. 2019). Thus, to investigate the influence of fresh verses heated samples in the mixing model we fit the model using both fresh verses heated sulfide samples independently. Standard errors for the estimated best-fit parameters were derived from 1,000 iterations that generated synthetic reflectance spectra of each surface unit centered around the mean and corresponding standard deviation. The optimal model solution was determined by minimizing the root mean square error (RMSE) between the observed and modeled spectra using NNLS in each iteration. The suite of plausible spectral fits for each cluster is represented by the mean and corresponding standard deviations of the best model fits calculated from all iterations. We computed the mean and standard deviation of the fractional abundance (Table 1) using the results of the best-fit parameters from both fresh and heated sulfide samples. Thus, possible each solution falls within the range defined by the mean ± 1σ standard error of the relative abundance for overlapping compositions. We evaluated the influence of fresh versus heated samples on the spectral modeling using the RMSEs and comparing their resulted chi-square ($\chi^2$) values for each surface unit. We validated the $\chi^2$ values for fresh and heated samples using F-test (e.g., Bevington and Robinson, 2002) at 95% confidence interval to verify if the difference in $\chi^2$ for fresh verses heated sample in spectral modeling is statistically significant.

For convenience, the fractional amount of augite and diopside is reported together as pyroxene minerals, and the fractional amount of labradorite and albite is reported together as plagioclase minerals. It is important to note that the percent mineral abundance estimated and reported here (Table 1) does not represent an absolute quantification of the minerals present on the surface, but rather a modeled contribution of laboratory mineral phases to the spectra of corresponding spectral classes identified within the MDIS-WAC observations.

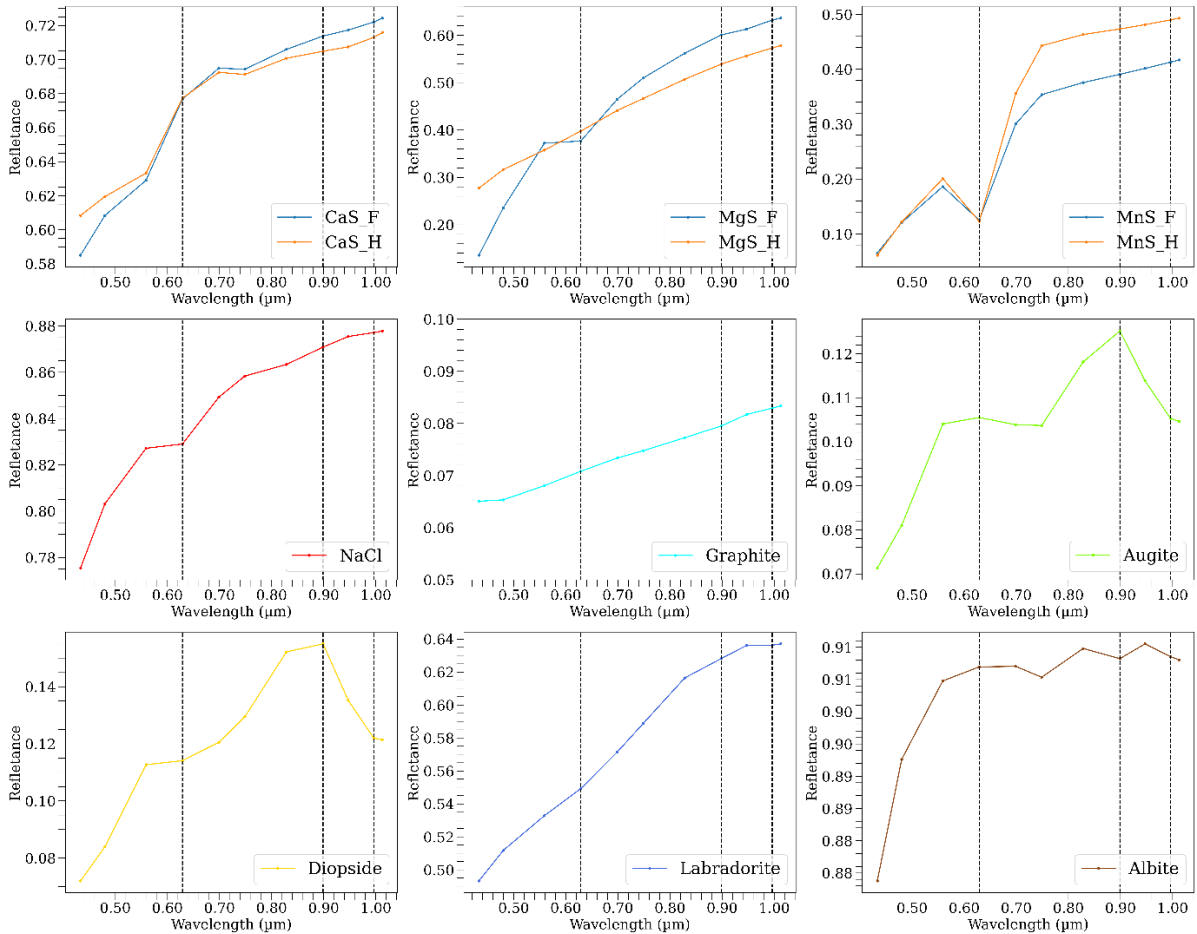

**Fig. 4**: Reflectance spectra for laboratory endmember CaS, MgS, MnS, NaCl, graphite, augite, diopside, labradorite, and albite resampled at MDIS wavelengths used for the linear spectral modeling. For sulfides, *_F and *_H represent fresh and heated samples, respectively. The black dashed vertical lines indicate characteristic absorption features at ~0.63, 0.9, and 1.0 μm. The reflectance spectra of sulfides were collected from Varatharajan et al. (2019), while the spectra of chlorides, graphite, and silicate minerals were collected from the RELAB database (Milliken et al. 2021).

## 3. Results

### *3.1. Surface units and spectral characteristics*

The seven distinct spectral clusters identified in this instance are shown in Fig. 2b, where C2 and C3 units represent the waning and active hollows at the crater, respectively. The spectra in Fig. 2d exhibit a broad and deep absorption feature around the 0.75 μm (with a shoulder at ~0.7 μm) with a similar strength across all units, with a maximum band depth of 5%. An absorption feature at 0.75 μm has also been reported in Mercury's global average spectra (Denevi et al. 2009), hollow floors and halos (Wang et al. 2020), hollows and pyroclastic deposits (Pajola et al. 2021; Lucchetti et al. 2021), and the high-reflectance plains without hollows (Murchie et al. 2015). This suggests that the absorption feature at 0.75 μm is less likely attributable to substantial compositional differences between the hollows and the surrounding terrains and is not related to volatiles phase within the hollows (Wang et al. 2020). The laboratory spectrum of augite, resampled at MDIS wavelengths (Fig. 4), reveals a broad absorption feature at 0.75 μm. Albite also exhibits an absorption feature at 0.75 μm, but it is much narrower than that observed in the MDIS spectra. Among sulfides, CaS (both fresh and heated) shows only a weak absorption at 0.75 μm. Therefore, we expect augite to be present in all surface units at the crater. Other absorption features, such as those near ~0.63 μm and between 0.9-1.0 μm, appear unique to specific surface units and may indicate real compositional differences between the spectral classes (Fig. 2). However, of these features, only the one near ~0.63 μm is relatively deep, with a band depth of ~2%, while those between 0.9-1.0 μm have a band depth of less than 1% (Fig. 2d). The relative abundances of these minerals (fractional contribution of endmember spectra) in each unit are estimated in the spectral modeling section (*Section 3.3*).

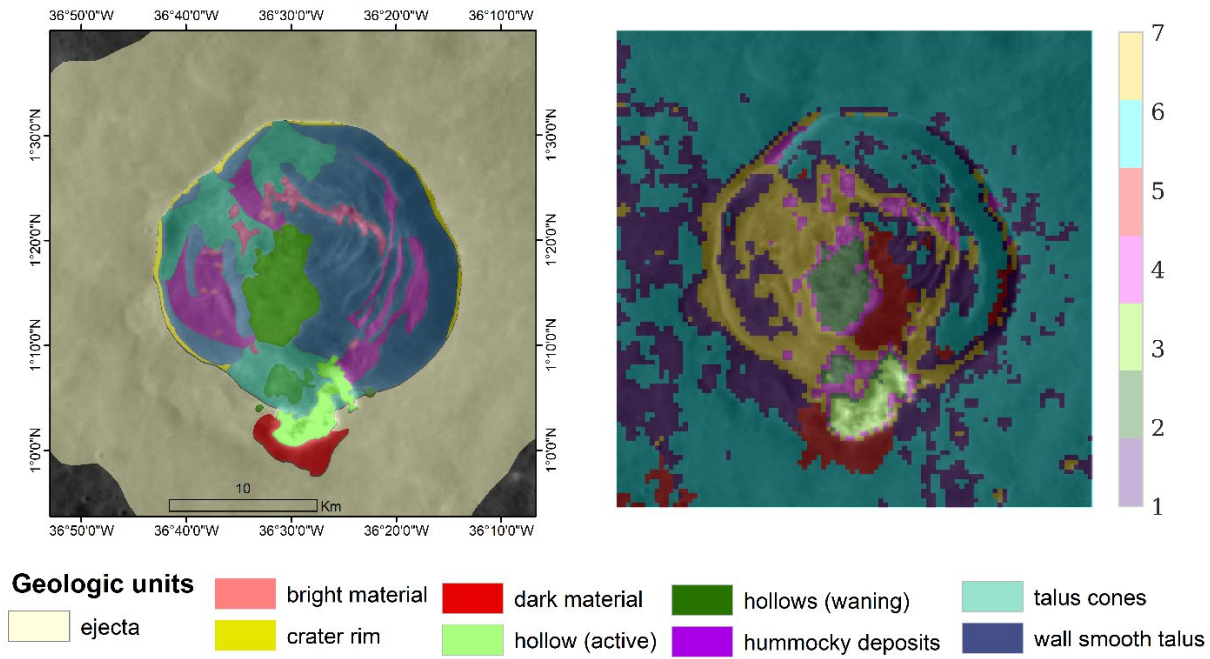

**Fig. 5**: (a) Geologic map of Dominici crater and surrounding areas adapted from Lucchetti et al. (2018) overlain on the MDIS basemap. The colors of geologic units correspond to colors and names of the units given in lower part of the figure. (b) Seven distinct spectral classes were identified in and around the crater. The colors and assigned names of the spectral classes are provided on the right side of the figure. North is up.

Spectral mapping using PC-GMM aligns well with geomorphic units within and around the crater (Fig. 5). Both C1 and C6 coincide broadly with crater ejecta as mapped by Lucchetti et al. (2018) using MDIS-NAC images (Fig. 5). The ejecta units exhibit a slight spectral difference between them but are otherwise lacking in spectral features aside from the broad absorption at 0.75 μm (Fig. 2). The probability plot at the MDIS-WAC pixel level further indicates that many pixels share probabilities across both units (Fig. A3), supporting slight differences between these spectral units. These units have steeper slopes in the normalized spectra at 0.558 μm, with C6 having the steepest one (Fig. 2c). The two families of ejecta in C1 and C6 may represent vertical heterogeneity in the crust, likely due to compositional differences (discussed in later section) between these ejecta units. These observations are consistent with the normalized average spectra for ejecta units discussed in Lucchetti et al. (2018).

The C7 unit mostly corresponds to Lucchetti et al.'s (2018) "talus cones" and some parts of the wall and rims (hereafter denote simply as talus unit) and likely represents slumps within the crater and along part of the crater rim (Fig. 5). C7 is distinguished from other spectral classes by its reflectance and the slope of the shoulder on the 0.75 µm band (Figs. 2c and 2d). A portion of the area mapped as C5 corresponds to the "dark material" (DM) unit mapped by Lucchetti et al. (2018), but C5 occurs elsewhere as well (Fig. 5). C5 is observed near both the active and waning hollows at the crater, although Lucchetti et al. (2018) found a DM unit only in proximity to the active hollow in the crater wall/rim. C5 shows no absorptions at 0.63 and 1.01 µm, but it is the only spectral unit that shows a very weak absorption at ~0.95-µm, with a band depth <1% (Fig. 2d). An absorption feature at ~0.95 µm in the laboratory spectra is not readily identifiable (Fig. 4). The slope of the normalized reflectance for C5 falls between that of crater ejecta and talus deposits (Fig. 2c).

The actively forming hollow in the southern wall/rim of the crater corresponds to C3, where the high-resolution basemap shows a sharp margin and fresh morphology (Fig. 5; Lucchetti et al. 2018). The normalized spectrum at 0.558 µm and the continuum-removed spectrum for C3 shows absorptions at ~0.63 and 1.0 µm (Fig. 2). Compared to other spectral units, C3 exhibits the strongest absorption at 0.63-µm and is the only spectral unit with a subtle absorption at ~1.0 µm. The normalized spectrum of this active hollow has the lowest slope among all spectral units (Fig. 2c)— consistent with the spectral characterization of Lucchetti et al. (2018). The waning stage hollows in the crater center and along the southern wall occur within C2 (Fig. 5). The normalized and continuum-removed spectra for C2 show a weaker absorption at 0.63 µm and lack a ~1.0 µm feature compared to the active hollow (Fig. 2). However, waning hollows show an absorption feature at ~0.9 µm that the active hollow lacks. The slope of the normalized spectra of the waning hollows is gentle but slightly steeper than that of the active hollow (e.g., Lucchetti et al. 2018). Among sulfides, MgS and MnS, as well as NaCl, show an absorption at 0.63 µm, while albite exhibits a weak absorption feature around ~0.9 µm (Fig. 4). The C4 unit corresponds to the edge of the waning hollow and a few other isolated spots inside the crater (notably coexisting with the talus deposits unit; Fig. 5) and does not exhibit ~0.63, 0.9, or 1.0-µm absorptions. The slope of the normalized spectra for this unit falls between those of the hollows and other units (Fig. 2c). The C4 unit appears to be transitional unit between the hollows and the surrounding terrains.

## 3.2. Spectral slopes

All spectral units in Dominici exhibit VISr 430/750 nm ratio values lower than 1 (Fig. 3). This implies that all spectral units have a positive or red spectral slope in visible wavelengths. This characteristic is consistent with the spectral investigation of hollows and their surrounding terrains elsewhere on the planets using both the MASCS data (e.g., Barraud et al. 2020, 2023) and MDIS channels (e.g., Blewett et al. 2013; Thomas et al. 2016). In contrast, all spectral units exhibit IRr 950/750 nm ratio values higher than 1 meaning that reflectance also increases with an increase in infrared wavelengths. While the ejecta units (C1 and C6) showed the lowest VISr ratio values, they showed the highest IRr ratio values. Conversely, both active and waning hollows (C3 and C2) showed the highest ratio values for VISr but the lowest ratio values for IRr. While the waning hollows show a flatter VISr spectral slope (430 nm/750 nm), the bright active hollow has the flattest slope but higher reflectance. The active hollow shows a shallower slope and is spectrally "bluer" compared to surrounding terrains at MDIS wavelengths (Blewett et al. 2011, 2013; Thomas et al. 2014, 2016; Lucchetti et al. 2018). The talus deposits and DM units have spectral ratio values at VISr and IRs comparable to the crater ejecta units. Lastly, C4 shows spectral ratio values between the hollows and other units, consistent with the characteristics of the slope of the normalized spectra for this unit.

## 3.3. Mineral abundance (fractional contribution)

Spectral modeling of all surface units (C1–C7) using fresh and heated sulfides independently is presented in Figs. 6 and 7, respectively. Using heated samples, we found relatively higher $\chi^2$ values for all surface units except for crater ejecta— which theoretically indicating that fresh sulfide samples might provide an overall better fit to MDIS-WAC observation. However, the $\chi^2$ differences are extremely small for the corresponding cluster, meaning that the models are almost indistinguishable. A comparison of $\chi^2$ values for the corresponding cluster at the 95% confidence interval using an F-test indicates that the differences in none of the surface units are statistically significant. Therefore, we conclude that the results in Table 1 are not significantly affected by the choice of fresh or heated sulfide endmembers in the modeling using MDIS-WAC observations. The mean and standard deviation of the fractional contribution of endmembers show either a complete absence or only trace amounts of MnS, NaCl, and labradorite minerals (Table 1). Therefore, we omit their abundances from further interpretation in this paper. The terrains surrounding Dominici crater, including units C5 and C7, are predominantly composed of graphite, pyroxene, and Mg- and Ca-sulfides. The mean abundances for C5 and C7 are 33% and 26% graphite, 39% and 45% pyroxene (augite), 11% and 10% MgS, and 9% and 10% CaS, respectively (Table 1). C1 and C6, the crater ejecta units, contain slightly higher amounts of graphite (36% and 37%, respectively) but slightly lower amounts of pyroxene (38% and 30%) compared to the surrounding terrains (C5 and C7). Both C1 and C6 shows elevated MgS content (15% and 21% means, respectively) than the surrounding units. The slightly redder spectra of C6 compared to the spectra of the C1 unit (as discussed in *Section 3.1*) might be due to the slightly elevated MgS content in the former unit compared to the latter. However, the CaS abundance in both ejecta units is lower than in C5 and C7. These spectral units (C1 and C5–C7) also show an almost absence of plagioclase minerals (labradorite and albite).

**Table 1**: Abundance of minerals (fractional contribution of endmember spectra in surface unit spectra) in percent (mean± 1-σ standard error) using best-fit parameters from the non-negative constrained least squares method spectral fitting for the surface units.

| Unit (designation) / Minerals | C1 (Ejecta unit) | C2 (Waning hollow) | C3 (Active hollow) | C4 (Transitional unit) | C5 (Dark material) | C6 (Ejecta unit) | C7 (Talus unit) |
|---|---|---|---|---|---|---|---|
| CaS | 7 ± 7 | 7 ± 10 | 3 ± 5 | 10 ± 14 | 9 ± 11 | 7 ± 9 | 10 ± 11 |
| MgS | 15 ± 7 | 1 ± 2 | 0 ± 1 | 5 ± 6 | 11 ± 8 | 21 ± 10 | 10 ± 7 |
| MnS | 0 ± 0 | 1 ± 2 | 0 ± 0 | 1 ± 3 | 0 ± 1 | 0 ± 1 | 0 ± 1 |
| NaCl | 0 ± 0 | 2 ± 6 | 1 ± 3 | 1 ± 5 | 0 ± 2 | 0 ± 2 | 1 ± 3 |
| Graphite | 36 ± 25 | 8 ± 17 | 5 ± 11 | 21 ± 28 | 33 ± 31 | 37 ± 29 | 26 ± 29 |
| Augite [a] | 38 ± 13 | 54 ± 25 | 59 ± 18 | 48 ± 25 | 39 ± 22 | 30 ± 19 | 45 ± 21 |
| Diopside [a] | 0 ± 1 | 9 ± 18 | 7 ± 14 | 4 ± 11 | 2 ± 8 | 2 ± 6 | 2 ± 6 |
| Labradorite [b] | 0 ± 0 | 0 ± 2 | 0 ± 1 | 1 ± 7 | 0 ± 3 | 0 ± 0 | 0 ± 4 |
| Albite [b] | 0 ± 0 | 15 ± 10 | 21 ± 6 | 4 ± 8 | 1 ± 3 | 0 ± 0 | 1 ± 4 |

*Note*: [a] indicates pyroxene and [b] indicates plagioclas minerals.

The active hollow (C3) spectrum is modeled by the lowest amounts of MgS (near zero), CaS (3%), and graphite (5%) among all units. However, it has the highest abundance of silicate minerals, including 59% augite, 21% albite, and 7% diopside. Conversely, the waning hollow spectrum (C2) is modeled to have a similar amount of MgS (1%) and slightly higher graphite (8%), but its CaS abundance is nearly double that of the active hollow. The silicate content modeled in C2—54% augite and 15% albite—is slightly lower than in C3. The edge of the waning hollow unit spectrum (C4) is modeled to contain relatively higher amounts of MgS (5%) and graphite (21%) compared to both active and waning hollows, while its silicate content—48% augite and 4% albite—is intermediate between the surrounding terrains and the hollows. C4 is modeled by the CaS abundance of 10%, similar to the C7.

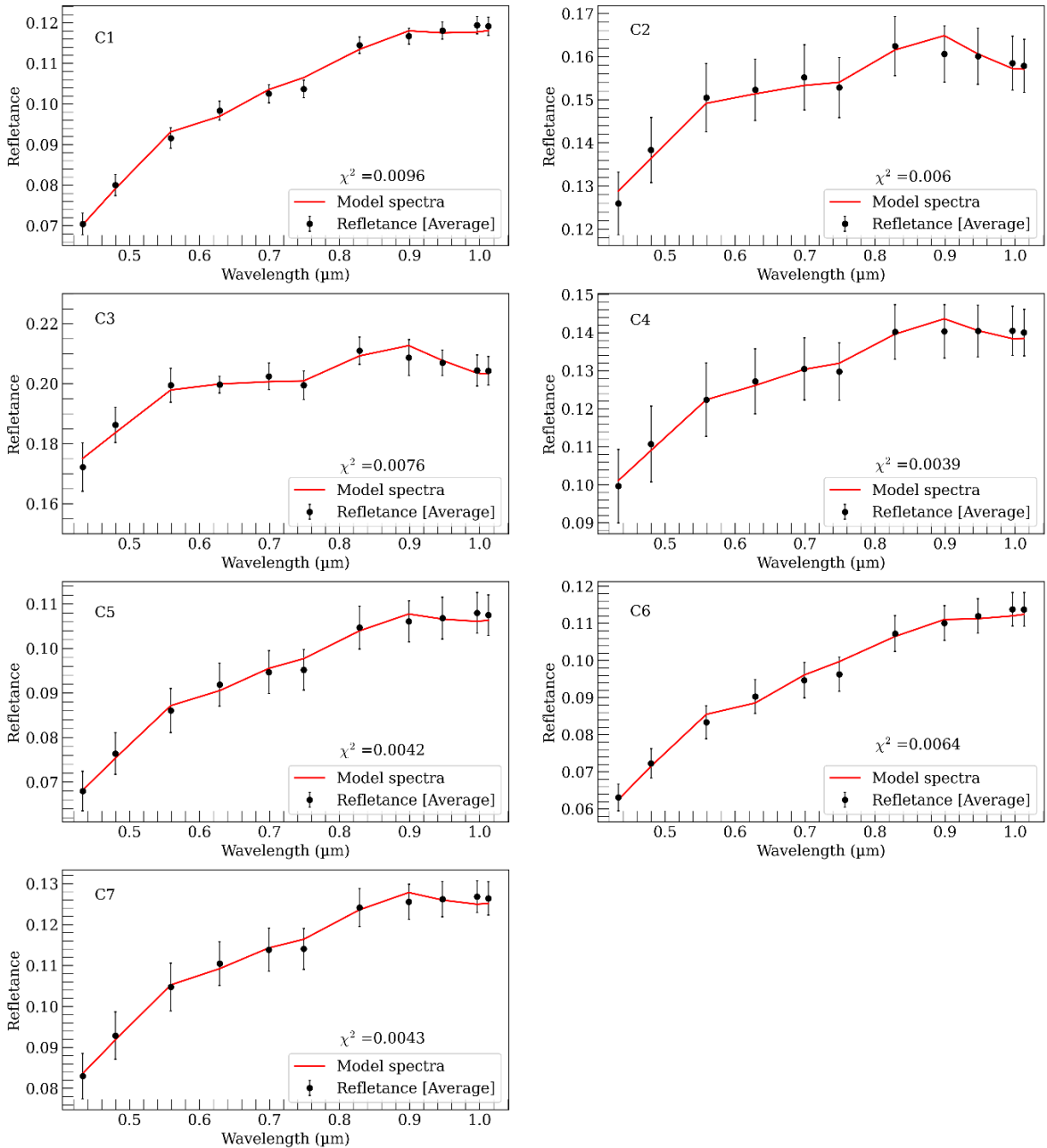

**Fig. 6**: Model fits using non-negative constrained least squares method spectral fitting for the surface units (C1-C7) using fresh sulfide samples. The average and 1-σ standard error reflectance spectra (black error bars) of the surface units and the best-fit parameters model spectra (red line) using the NNLS approach. The modeled $\chi^2$ value for the corresponding surface units is provided in the subplots.

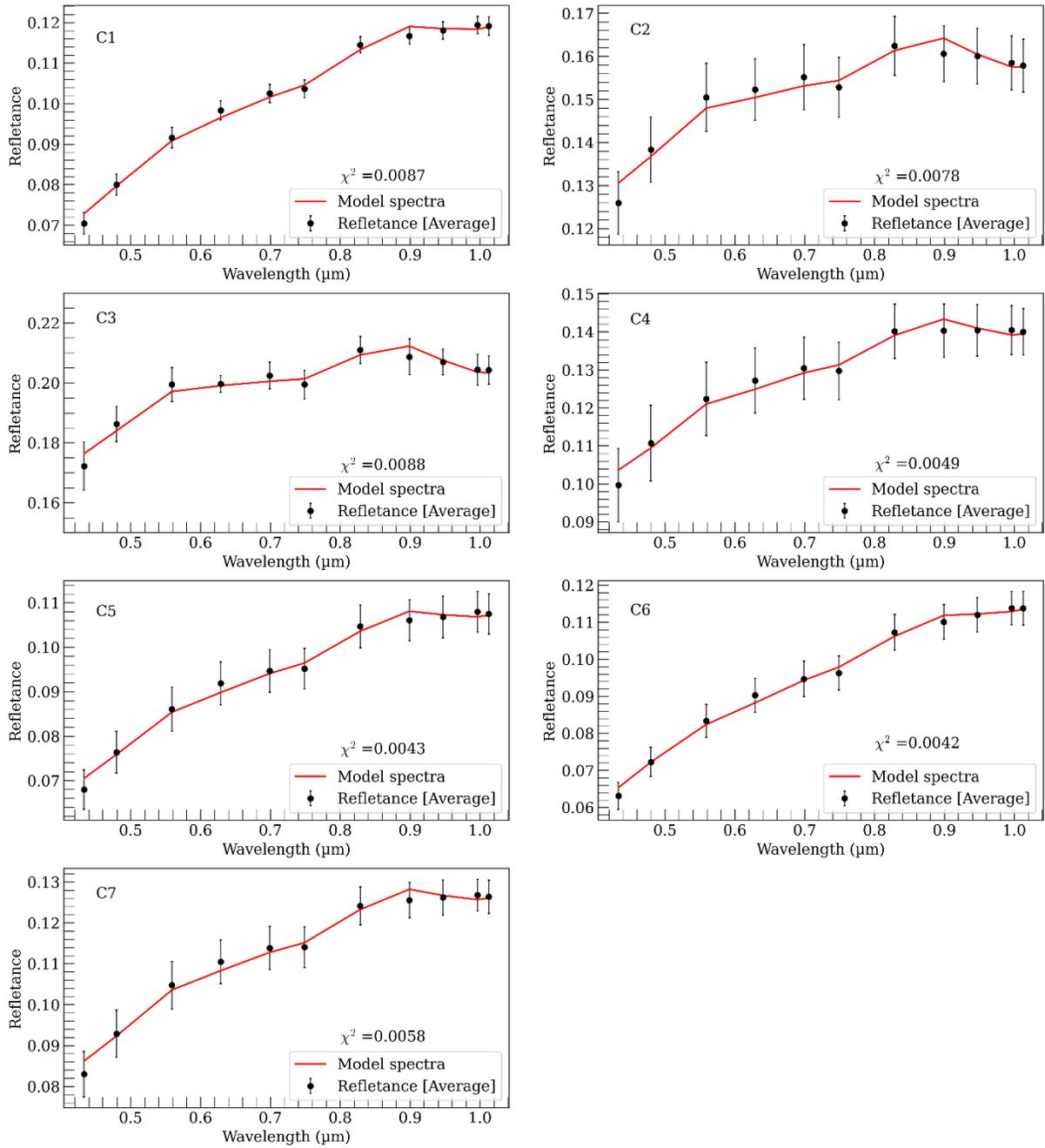

**Fig. 7**: Model fits using non-negative constrained least squares method spectral fitting for the surface units (C1-C7) using heated sulfide samples. The average and 1-σ standard error reflectance spectra (black error bars) of the surface units and the best-fit parameters model spectra (red line) using the NNLS approach. The modeled $\chi^2$ value for the corresponding surface units is provided in the subplots.

## 4. Discussion

The application of machine learning in this study has enabled high fidelity mapping of the hollows in Dominici crater at different developmental and devolatilization phases including both the active and waning stages. The spatial boundaries of distinct spectral classes closely follow geologic and geomorphic contacts within the area, as visually observed in higher resolution images (Figs. 2 and 5). Given the visible surface geomorphology and spectral characteristics from MDIS data (see *Section 3*), we consider the DM (C5) and talus deposits (C7) units to be similar to the LRM surrounding Dominici. The ejecta units (C1 and C6) represent materials that have been excavated and redistributed onto the surface by the Dominici impact into target LRM. However, the compositions of ejecta still represent the LRM as their modeled composition is largely consistent with that of the talus and DM, with a higher modeled amount of graphite and sulfides. The relatively darker appearance of these terrains in LRM is consistent with the abundance of graphite and sulfide, consistent with existing literature (e.g., Blewett et al. 2013; Xiao et al. 2013; Murchie et al. 2015; Weider et al. 2015). This further supports the presence of carbon in the form of graphite in LRM, as observed by various instruments aboard the MESSENGER spacecraft (e.g., Peplowski et al. 2016; Klima et al. 2018). The relatively elevated MgS content modeled in ejecta could be explained by the impact event exposing Mg-sulfide on the surface. Thus, sulfides (MgS and CaS) are potentially the primary volatile-bearing species for LRM in the area. Our results align with previous findings that attribute LRM to enrichment in Mg, Ca, and S, with an abundance of high-reflectance sulfides as volatile-bearing species (e.g., Thomas et al. 2016).

An absorption feature at ~0.63 μm for hollow spectra suggests the presence of sulfides (Vilas et al. 2016; Lucchetti et al. 2018) or chlorides (Vaughan et al. 2012; Lucchetti et al. 2021), in addition to graphite (e.g., Murchie et al. 2015; Wang et al. 2020). The weakness of this absorption feature in hollows spectra might be due to the extreme thermal cycling that can result in weakening the spectral absorption features of sulfides (Helbert et al. 2012, 2013). Although we found a ~0.63 μm absorption band exclusively in the hollow spectra (Fig. 2), spectral modeling revealed that hollow materials show the lowest combined abundance of sulfides (MgS and CaS). This can be explained by the lower graphite content in the hollows, which can otherwise mask the ~0.63 μm absorption band at WAC wavelengths (Fig. 4). Thus, the absence of the ~0.63 μm absorption band in other spectral units does not necessarily indicate the absence of sulfides in those units. Nevertheless, the higher VISr band ratio values of the hollows compared to surrounding terrains are largely consistent with the results from other hollows on Mercury (e.g., Blewett et al. 2011, 2013; Thomas et al. 2014, 2016). For instance, the visible spectral slope (VISr of 415 nm /750 nm) from MASCS data exhibits, in general, the lowest value in LRM and crater floor (Izenberg et al. 2014, 2015; Thomas et al. 2016). Moreover, VISr from MDIS data for different sites further confirms the lowest value in LRM and crater wall/floor materials (Thomas et al. 2016)—corresponding to ejecta, talus, and DM units in this study (see Fig. 5). The talus deposits and DM units have spectral ratio values at VISr and IRs comparable to the crater ejecta units.

On Mercury, volatile elements include sulfur, carbon (C), and chlorine (Cl), with varying weight percentages (wt%) for each species (e.g., Nittler et al. 2011; Weider et al. 2015; Peplowski et al. 2011, 2016; Evans et al. 2015). The materials responsible hollow formation and subsequently limiting further vertical growth can be inferred from their reflectance spectra. This is because hollows at the crater may not be entirely depleted of volatiles-bearing species and likely contain chemical phases involved in the formation process (e.g., Wang et al. 2020; Barraud et al. 2023). The candidate species for hollow formation are Ca- and Mg-sulfides, supported by several factors, such as the spectral characteristics of hollows (e.g., Blewett et al. 2011, 2013; Helbert et al. 2013), the thermophysical properties of elemental sulfur (Phillips et al. 2021), an elevated S concentrations on Mercury (e.g., Nittler et al. 2011, 2020; Evans et al. 2012), and the widespread presence of magnesium on the planet's surface and its correlation with sulfur (e.g., Stockstill-Cahill et al. 2012; Weider et al. 2015). Our results confirm the presence of sulfides (MgS and CaS) in the hollow spectra, in addition to the abundance of silicates. Laboratory spectra of mixtures of CaS and MgS (e.g., Helbert et al. 2013; Varatharajan et al. 2019) exhibit similar spectral characteristics (Vilas et al. 2016; Lucchetti et al. 2018) to those observed in the hollow spectra at Dominici. Experimental study indicates that sulfides (MgS and CaS) begin to decompose below 500°C in a vacuum environment (far below their melting temperature)—comparable to Mercury's daytime temperature (Helbert et al. 2013). Thus, thermally unstable MgS and CaS (e.g., Helbert et al. 2013; Wang et al. 2022) that produce S are the most likely volatile phase to hollow formation at the crater. Thermophysical models further support that sulfur's volatile behavior best represents the characteristics of the hollow-forming phase observed on Mercury's surface (Phillips et al. 2021).

Besides sulfides, other candidate volatiles-bearing species such as chloride minerals have also been proposed as a potential source material for hollows formation on Mercury (e.g., Vaughan et al. 2012; Lucchetti et al. 2021; Rodriguez et al. 2023). Although both heated and fresh laboratory MnS (e.g., Helbert et al. 2013; Varatharajan et al. 2019) and NaCl show an absorption at ~0.63 μm (Fig. 4), the spectral modeling results (Table 1) in this study show no substantial fractional contribution of MnS and NaCl in the hollow spectra. A minimal fractional abundance of MnS (Table 1) is consistent with the fact that the mineral has a very low upper limit abundance on Mercury (Peplowski et al. 2012). Thus, a lower contribution of MnS in the spectral modeling is expected since the Mercurian surface shows a lower concentration of Mn (e.g., Boukaré et al., 2019). On the other hand, linear mixture modeling of hollow spectra using MASCS data confirms a lower match with NaCl than sulfides (Barraud et al. 2023). This is supported by the fact that the wt% of Cl on Mercury's surface is much lower compared to S or C, with Cl content reaching at most ~0.2 wt% (Evans et al. 2015), while S and C can reach up to ~4 wt% (Nittler et al. 2011; Weider et al. 2015; Peplowski et al. 2011, 2016). The presence of weak absorption bands around ~0.63, 0.9, and 1.0 μm in the hollow spectra (Fig. 2) suggests a lower influence of graphite in the hollows. Further, our spectral modeling shows the least contribution of graphite to the hollow spectra, supporting the possibility that carbon has been lost by sputtering and space weathering during the development of the hollow (e.g., Vilas et al. 2016; Murchie et al. 2015; Wang et al. 2020; Blewett et al. 2018). The loss of the darkening agent graphite through the sputtering of carbon has also been proposed as a mechanism for hollow formation, leading to the brightening of the material (Blewett et al. 2018). On Mercury's surface, the destabilization of graphite in the planet's harsh environment may contribute to hollow formation (Blewett et al. 2016; Wang et al. 2020, 2022). The presence of a trace amount of graphite and sulfides in the waning hollow spectra might be due to the partial loss of volatile phase in LRM (Blewett et al. 2013). Another possibility is that devolatilization is not fully complete in the waning hollow and will continue until a complete loss of volatile species. Alternatively, the waning hollow may consist of larger particles and/or a rougher texture compared to the surrounding terrain, which could inhibit the complete depletion of the volatile species and left some of volatile-bearing phases as lag deposit (Thomas et al. 2016).

Previous studies report a broad absorption feature extending between 0.558 and 0.82 μm in hollows with different degrees of strength at the crater (Vilas et al. 2016; Lucchetti et al. 2018). We identified two subtle absorption features at ~0.90 and 1.0 μm in the hollows at Dominici Crater (Fig. 2), which may contribute to their distinct composition compared to other units, as revealed through spectral unmixing. Spectral modeling of hollows shows an elevated contribution of silicates, mostly pyroxene minerals. Hollows develop in silicate-crust materials and, thus, it is plausible that hollow spectra retain characteristics of Mercury's volcanic crust-forming silicate minerals (e.g., Weider et al. 2016; Namur and Charlier, 2017). Our spectral modeling reveals varying levels of augite (pyroxene) abundance across all surface units (Table 1), supported by the presence of the 0.75 μm band in all units (Fig. 2) and the laboratory augite spectra (Fig. 4). Spectral modeling of different hollows using the observations from the MASCS instrument has also found a match to a mixture of minerals including silicates (Barraud et al. 2023). Lucchetti et al. (2018) report bedrock-forming materials, such as pyroxene, to be partially responsible (besides sulfides) for the spectral absorption of hollows at different craters, including those at Dominici. Non-volatile phase, primarily silicate minerals such as pyroxene, might remain behind as lags after devolatilization leading to further restriction of the vertical growth of hollows at the crater. Moreover, a continuation of devolatilization would be prevalent in the absence of the lag deposits which would result in a complete loss of all volatile materials (e.g., Thomas et al. 2016). The presence of some trace amount of volatiles-bearing species with abundant non-volatiles silicates in the waning hollows supports the hypothesis that silicates are primarily left behind as lag deposits after partial devolatilization and might prevent the complete loss of volatile phases.

The spectral modeling and mineral mapping of this study support a possible evolution of Dominici's hollows as illustrated in Fig. 8. The LRM, abundant with graphite and sulfide throughout, is responsible for the darker appearance of the local region and serves as the background terrain and parent (source) material of future hollows. The Dominici impact event exposed volatile-bearing species (e.g., sulfides) at the surface in crater ejecta, with concentration of volatiles (sulfur and/or carbon) through impact melting and subsequent differentiation (Vaughan et al. 2012) or migrating melt assimilation (Helbert et al. 2012). Active formation of hollow begun with thermal desorption or decomposition of sulfides and sublimation loss of the concentrated volatiles at the surface or shallow subsurface. Thus, hollow formation proceed with the devolatilization of primarily sulfur, with carbon loss from sputtering by space weathering or photon bombardment (e.g., Blewett et al. 2016, 2018; Wang et al. 2020). As the devolatilization of sulfur and carbon ceases, the hollows move into the waning stage, leaving behind lag deposits consisting primarily of a mixture of non-volatile silicate minerals and some volatiles-bearing species where devolatilization is incomplete. These mixed lags are likely responsible for halting further development and deepening of the hollows. As volatile loss continues, a higher silicate contribution is seen in the spectral modeling. This might suggest that a lower amount of sulfides left in the hollow restricts further growth of hollows— consistent with the results using spectral modeling of MASCS-VIRS data (Barraud et al. 2023).

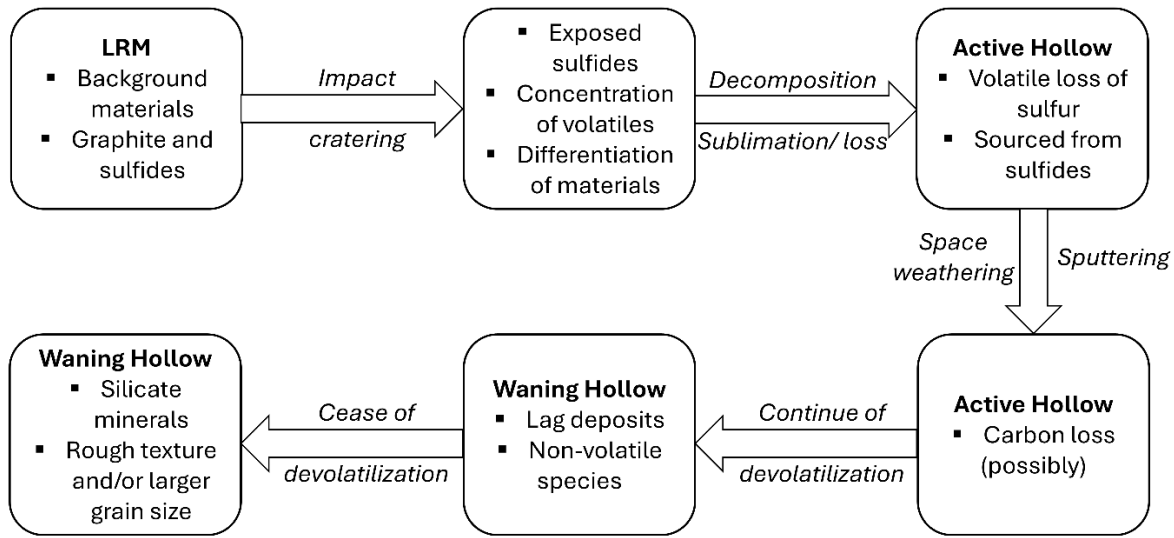

**Fig. 8:** A simplified graphical illustration of the possible formation and evolution of hollows at the crater. Rectangular boxes characterize surface units and their corresponding compositions, while the arrows represent the process involved in transitioning from one phase to another.

We emphasize that hollows on Mercury's surface are not exclusive to LRM (e.g., Bickel et al. 2025; Deutsch et al. 2025), and the mineralogy of Mercury surface is diverse (e.g., Namur and Charlier, 2017; McCoy et al. 2018; Nittler and Weider, 2019) and, thus, the hollow evolution outlined here represents their development within LRM at Dominici crater. However, a viability of the formation model may also be plausible for hollows that occurs outside LRM such as high reflectance smooth plains (HRP) or other regions of the planet. The spectral characteristics and volatile phases associated with hollow formation vary depending on the local geochemistry of the substrate (e.g., Vaughan et al. 2012; Lucchetti et al. 2018, 2021; Barraud et al. 2020; Pajola et al. 2021). While sulfides are the primary volatiles-bearing species contributing to hollow formation at the crater, the potential role of chloride minerals as volatile-bearing species cannot be ruled out for the hollows elsewhere on Mercury (Vaughan et al. 2012; Lucchetti et al. 2021; Rodriguez et al. 2023). Recently, Deutsch et al. (2025) conducted a global-scale analysis of hollow degradation states using the global hollow database (Bickel et al. 2025) and classified hollows into three degradation stages. Hollows with sharp morphology and high visible reflectance interiors were assigned to 'Stage 1', while those lacking high visible reflectance but retaining distinct morphology were categorized as 'Stage 2' (Deutsch et al. 2025). The hollows examined in this paper are an excellent example of the variability in 'Stage 1' hollows as described by Deutsch et al. (2025), suggesting that the development of hollows may be a bit more complex than that shown to date.

## 5. Conclusion

Extraction of reflectance spectra of hollows at different devolatilization phases and their composition analysis is key to understanding the development of hollows, characteristics of volatiles and non-volatiles, and the composition of low reflectance materials on Mercury. This study leverages PC-GMM (Emran et al. 2023a) and linear spectral modeling to provide insight into the formation and compositional evolution of hollows during different devolatilization phases. Lower reflectance materials exhibit a darker appearance in orbital observations due to the abundance of graphite (e.g., Murchie et al. 2015). The LRM also contains sulfides. The impact event that created Dominici crater further exposed and concentrated these volatiles via impact melting and subsequent differentiation of composition (Vaughan et al. 2012) or sulfur accumulation by migrating melts (Helbert et al. 2012). The thermal decomposition of sulfides and sublimation of concentrated volatiles leads to hollow formation, with continued loss of carbon via sputtering, space weathering, or photon bombardment (Blewett et al. 2016). As devolatilization continues, the hollows retain a lower amount of sulfides and leave a mixture of silicates as lag deposits. These primarily non-volatiles are responsible for restraining the further growth of the hollows. Future higher resolution spectroscopic observations from the visible–near-infrared imaging spectrometer (VIHI; Flamini et al. 2010; Cremonese et al. 2020) onboard ESA/JAXA's BepiColombo spacecraft mission (Benkhoff et al. 2010) to Mercury will help to further constrain our understanding of the formation of hollows and the volatile inventory of the planet.

**Acknowledgment**

This research was carried out at the Jet Propulsion Laboratory (JPL), California Institute of Technology, under a contract with the National Aeronautics and Space Administration (80NM0018D0004). We acknowledge JPL's High-Performance Computing (HPC) supercomputer facility, which was funded by JPL's Information and Technology Solutions Directorate. We also acknowledge Ariel Deutsch, David Blewett, Scott Murchie, and Laura Kerber for their suggestions during the initial processing of data. We acknowledge the MDIS/MESSENGER spacecraft mission and PDS Geoscience Node teams for collecting, storing, and disseminating the image data. The authors acknowledge Deborah Domingue and an anonymous reviewer for comments that improved this manuscript.

**Data Availability Statement**

All data used in this study can be found in the National Aeronautics and Space Administration's Planetary Data System (PDS). MDIS data used in this study can be directly accessed to PDS Geoscience Node's Mercury Data Explorer (https://ode.rsl.wustl.edu/mercury/). Laboratory spectral data used in this study was acquired by the Reflectance Experiment Laboratory (RELAB) available through the PDS Geosciences Node Spectral Library (https://pds-speclib.rsl.wustl.edu/). Laboratory spectral data for sulfides can be access at Varatharajan et al. (2019).

**Appendix**

**A1.** The MDIS-WAC filters used in this study.

**Table A1**: The details of the MDIS-WAC filters used in this study.

| EDR ID (Fiter ID) | Wavelengths (µm) | Bandwidth (µm) |
|---|---|---|
| EW1020550593C | 0.4799 | 0.009 |
| EW1020550597D | 0.5589 | 0.005 |
| EW1020550601E | 0.6288 | 0.004 |
| EW1020550605F | 0.4332 | 0.018 |
| EW1020550609G | 0.7487 | 0.005 |
| EW1020550613H | 0.947 | 0.005 |
| EW1020550617I | 0.9962 | 0.012 |
| EW1020550621J | 0.8988 | 0.004 |
| EW1020550625K | 1.0126 | 0.020 |
| EW1020550629L | 0.8284 | 0.004 |
| EW1020550633A | 0.6988 | 0.004 |

## A2: Machine Learning (PC-GMM) Approach

A subset of the calibrated MDIS-WAC mosaic (see *Section 2.1*) was cropped to include the Dominici crater and its surrounding areas, as shown in Fig. 1c. We reduced the dimensionality of the dataset using principal component analysis (PCA; Pearson, 1901) to minimize the influence of possible noise. The hyperspectral signal identification by minimum error algorithm (*HySime*; Bioucas-Dias & Nascimento, 2008) was applied to automatically select the optimal number of principal component (PC-axes; Abdi & Williams, 2010) for the subset of MDIS-WAC data. Based on *HySime* result, we selected first two PC-axes for this study— explaining more than 99% of the variance in the original MDIS-WAC data. For reference, the first four PC-axes from the MDIS-WAC data for the Dominici crater and its surroundings are provided in Fig. A1.

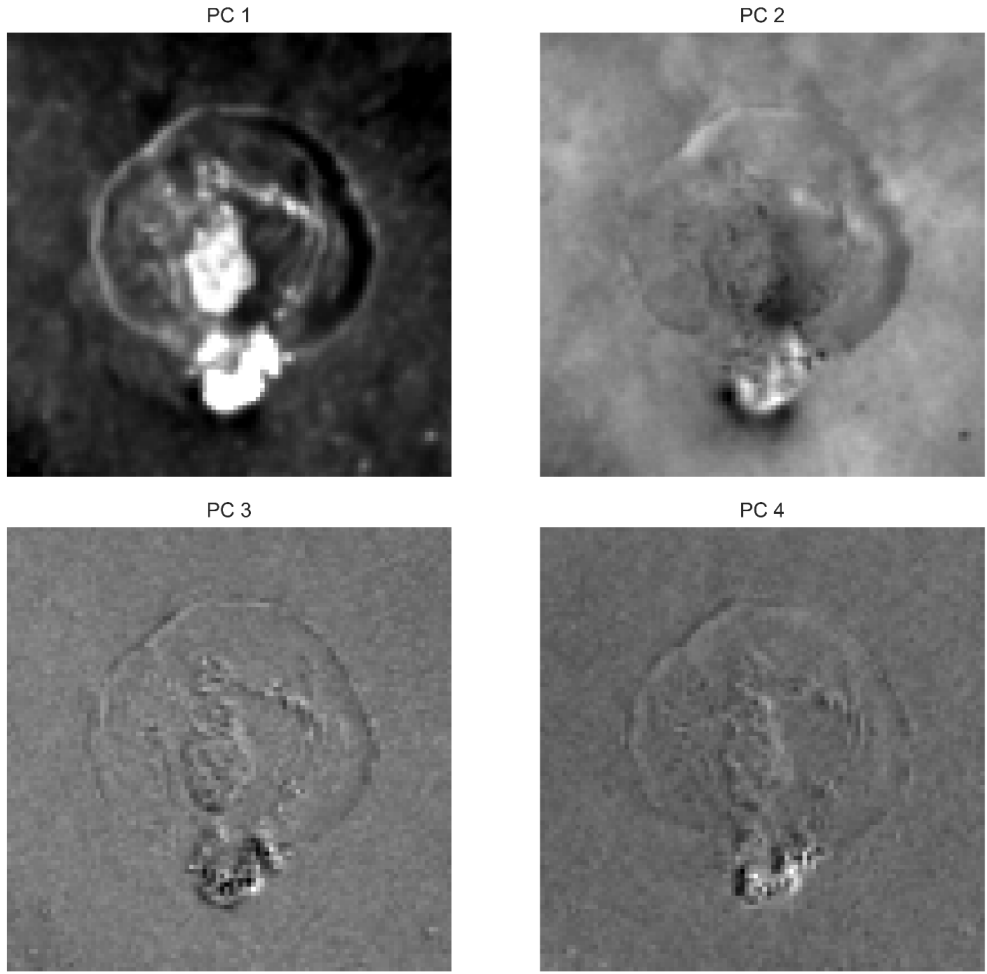

**Fig. A1.** The first four principal components from the subset of MDIS-WAC data at Dominici crater and surroundings.

After computing the PC-axes, we applied the multivariate Gaussian mixture model (GMM; Berge & Schistad Solberg, 2006; Li et al. 2014) to the first two PC-axes. The optimal number of clusters (#N) for GMM was determined using the Akaike information criterion (AIC; Akaike, 1974) and the Bayesian information criterion (BIC; Schwarz, 1978). Conventionally, a lower AIC value indicates a better model fit to the data (e.g., Liddle, 2007), while the optimal number of clusters using BIC corresponds to the first local minimum of BIC values (e.g., Dasgupta & Raftery, 1998; Fraley & Raftery, 1998). Based on the AIC and BIC criteria (Fig. A2), we selected seven as the optimal number of clusters for GMM in this study.

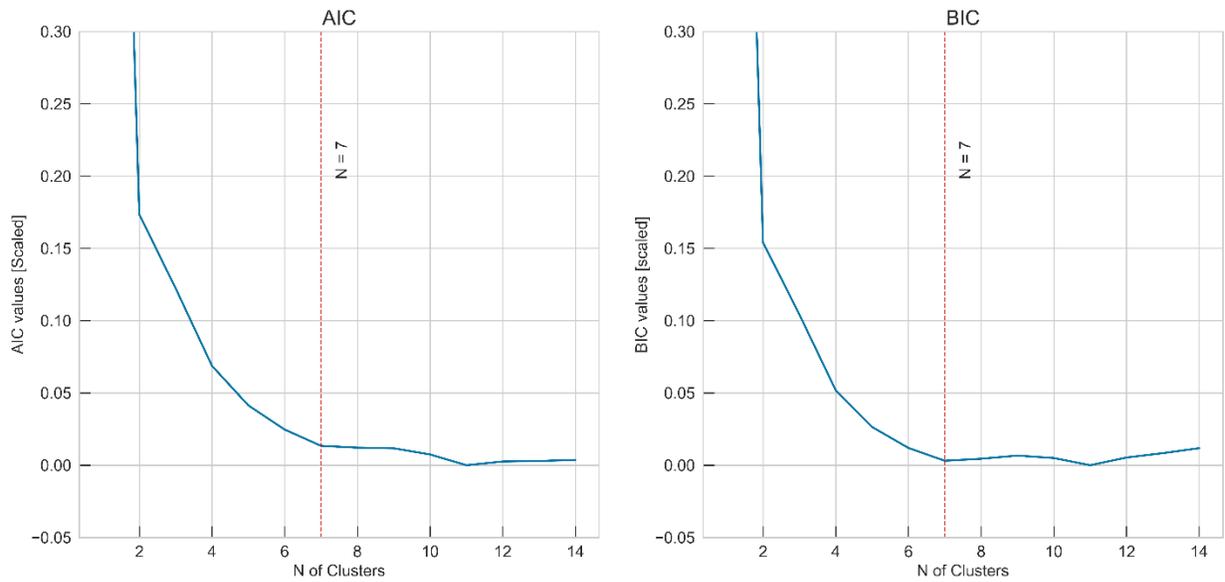

**Fig. A2.** Scaled AIC and BIC values at the different number of clusters for GMM at Dominici and surrounding areas.

**A3.** Likelihood of each surface unit at MDIS-WAC image pixel level using the PC-GMM.

The unsupervised machine learning, PC-GMM, used in this study uses a probability function approach that allows estimation of the likelihood of each spectral unit at the MDIS-WAC pixel level.

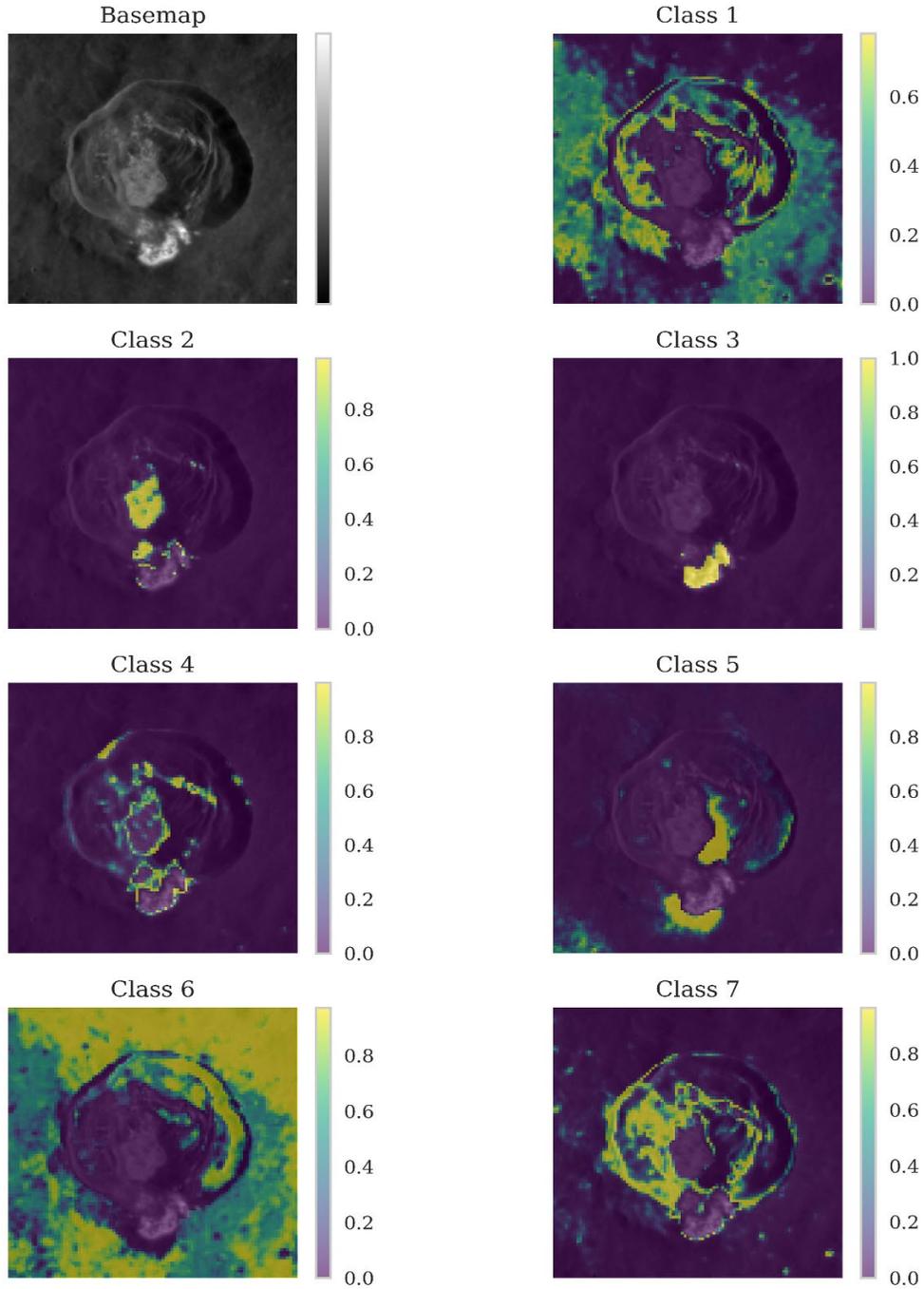

**Fig. A3**: The probability plot of each surface unit at the MDIS-WAC image pixel level using the PC-GMM. The subplots are superposed on the MDIS basemap. The probability scale ranges from 0 to 1, where higher values indicate a higher probability of the corresponding unit.

**A4.** The average and 1-σ standard error reflectance spectra of the surface units.

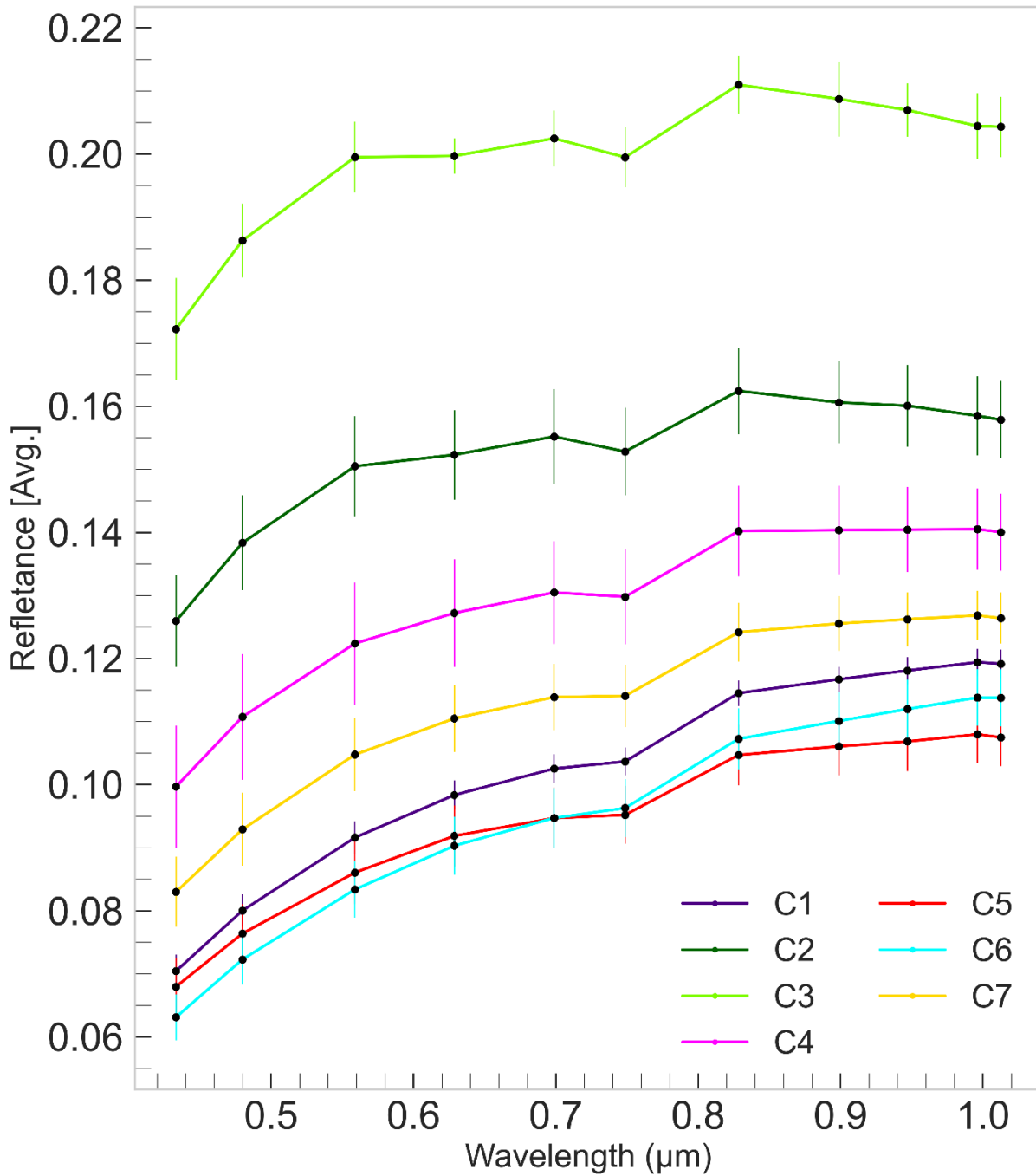

**Fig. A4**: The average and 1-σ standard error reflectance spectra of the surface units identified at the crater and surrounding terrain. The colors of the spectra correspond to the colors in the cluster map in Fig. 2b.

**A5.** Laboratory endmember spectra used in this study.

**Table A2**: The details of the reflectance spectra from RELAB used in this study.

| Endmember name | Spectral ID | Grain Size (µm) | Wavelength (µm) |
|---|---|---|---|
| $MgCl_2$ | C1JB32A | 0 - 250 | 0.30 – 2.55 |
| $CaCl_2$ | C1JBG30A | 0 - 250 | 0.30 – 2.55 |
| NaCl | C1JBG60A | 0 - 250 | 0.30 - 2.55 |
| Graphite | CASC79 | 0 - 45 | 0.30 - 2.60 |
| Amorphous graphite | CASC80 | 0 - 45 | 0.30 – 2.60 |
| Graphite powder | C1JBA05 | 0 - 45 | 0.28 – 2.60 |
| Enstatite | C1PP43 | 45 - 90 | 0.30 - 2.60 |
| Augite | C1PP18 | 45 - 90 | 0.30 -2.60 |
| Diopside | C1PP29 | 45 - 90 | 0.30 – 2.60 |
| Labradorite | C1DH12 | 0 - 25 | 0.30 -2.60 |
| Albite | CAPA05 | 0 - 25 | 0.30 -2.60 |